\documentclass[twocolumn,trackchanges, twocolappendix] {aastex631}
\shorttitle{A high-velocity prominence eruption on V1355 Orionis}
\shortauthors{Inoue et al.}
\graphicspath{{./}{figures/}}
\usepackage{threeparttable}
\usepackage{color}
\usepackage{enumerate}

\begin{document}

\title{Detection of a high-velocity prominence eruption leading to a CME associated with a superflare \\
on the RS CVn-type star V1355 Orionis}

\author[0000-0003-3085-304X]{Shun Inoue}
\affiliation{ Department of Physics, Kyoto University, Kitashirakawa-Oiwake-cho, Sakyo-ku, Kyoto, 606-8502, Japan}
\email{inoue.shun.57c@kyoto-u.jp}

\author[0000-0003-0332-0811]{Hiroyuki Maehara}
\affiliation{Okayama Branch Office, Subaru Telescope, NAOJ, NINS, Kamogata, Asakuchi, Okayama, 719-0232, Japan}

\author[0000-0002-0412-0849]{Yuta Notsu}
\affiliation{Laboratory for Atmospheric and Space Physics, University of Colorado Boulder, 3665 Discovery Drive, Boulder, CO 80303, USA}
\affiliation{ National Solar Observatory, 3665 Discovery Drive, Boulder, CO 80303, USA}
\affiliation{Department of Earth and Planetary Sciences, Tokyo Institute of Technology, Ookayama, Meguro-ku, Tokyo, 152-8551, Japan}

\author[0000-0002-1297-9485]{Kosuke Namekata}
\affiliation{ALMA Project, NAOJ, NINS, Osawa, Mitaka, Tokyo, 181-8588, Japan}

\author[0000-0001-6653-8741]{Satoshi Honda}
\affiliation{Nishi-Harima Astronomical Observatory, Center for Astronomy, University of Hyogo, Sayo, Hyogo, 679-5313, Japan}

\author{Keiichi Namizaki}
\affiliation{Department of Astronomy, Kyoto University, Kitashirakawa-Oiwake-cho, Sakyo-ku, Kyoto, 606-8502, Japan}

\author{Daisaku Nogami}
\affiliation{Department of Astronomy, Kyoto University, Kitashirakawa-Oiwake-cho, Sakyo-ku, Kyoto, 606-8502, Japan}
\affiliation{Astronomical Observatory, Kyoto University, Sakyo-ku, Kyoto, 606-8502, Japan}

\author{Kazunari Shibata}
\affiliation{Kwasan Observatory, Kyoto University, Yamashina, Kyoto, 607-8471, Japan}
\affiliation{ School of Science and Engineering, Doshisha University, Kyotanabe, Kyoto, 610-0321, Japan}



\begin{abstract}
Stellar coronal mass ejections (CMEs) have recently received much attention for their impacts on exoplanets and stellar evolution.
Detecting prominence eruptions, the initial phase of CMEs, as the blue-shifted excess component of Balmer lines is a technique to capture stellar CMEs.
However, most of prominence eruptions identified thus far have been slow and less than the surface escape velocity. 
Therefore, whether these eruptions were developing into CMEs remained unknown.
In this study, we conducted simultaneous optical photometric observations with \textit{Transiting Exoplanet Survey Satellite} and optical spectroscopic observations with the 3.8m Seimei Telescope for the RS CVn-type star V1355 Orionis that frequently produces large-scale superflares.
We detected a superflare releasing $7.0 \times 10^{35} \: \mathrm{erg}$. In the early stage of this flare, a blue-shifted excess component of $\mathrm{H \alpha}$ extending its velocity up to $760-1690 \: \mathrm{km \: s^{-1}}$ was observed and thought to originate from prominence eruptions. 
The velocity greatly exceeds the escape velocity (i.e., $\sim 350 \: \mathrm{km \: s^{-1}}$), which provides important evidence that stellar prominence eruptions can develop into CMEs.
Furthermore, we found that the prominence is very massive ($9.5 \times 10^{18} \: \mathrm{g} < M < 1.4 \times 10^{21} \: \mathrm{g}$).
These data will clarify whether such events follow existing theories and scaling laws on solar flares and CMEs even when the energy scale far exceeds solar cases.
\end{abstract}



\section{Introduction} \label{sec:intro}
\begin{deluxetable*}{ccccccccccc}
\tablenum{1}
\tablecaption{Basic physical parameters of the K-type subgiant star of the binary \label{tab:V1355}}
\tablewidth{0pt}
\tablehead{
\colhead{Spectral Type} & \colhead{$V$\tablenotemark{$*$}}  & \colhead{$V - R_{C}$\tablenotemark{$*$}} & \colhead{$d$}\tablenotemark{$\dag$} & \colhead{$P_{\mathrm{orb}}$}\tablenotemark{$\ddag$} & \colhead{$P_{\mathrm{rot}}$}\tablenotemark{$\$$}
& \colhead{$R$}\tablenotemark{$\S$} & \colhead{$L$}\tablenotemark{$\P$} & \colhead{$M$}\tablenotemark{$\natural$} & \colhead{$T_\mathrm{eff}$}\tablenotemark{$\sharp$} &
\colhead{$v_e$} \tablenotemark{$\star$}
\\
\colhead{} & \colhead{(mag)}  & \colhead{(mag)} &
 \colhead{(pc)} & \colhead{(days)} & \colhead{(days)} & \colhead{($R_{\odot}$)} &
 \colhead{($L_{\odot}$)} & \colhead{($M_{\odot}$)} & \colhead{(K)} &
 \colhead{(km/s)}
}
\decimalcolnumbers
\startdata
K0-2IV & 8.98 & 0.53 & 127.4 & 3.87 & 3.86 & 4.1 & 6.4
& 1.3 & 4750 & -347
\\
\enddata
\tablenotetext{$*$}{The $V$-band magnitude and difference between $V$ and and $R_{C}$-bands.}
\tablenotetext{$\dag$}{Stellar distance.}
\tablenotetext{$\ddag$}{Orbital period.}
\tablenotetext{$\$$}{Rotation period.}
\tablenotetext{$\S$}{Radius.}
\tablenotetext{$\P$}{Luminosity.}
\tablenotetext{\natural}{Mass.}
\tablenotetext{\sharp}{Effective temperature.}
\tablenotetext{\star}{Escape velocity at the surface.}
\tablerefs{(1),(5)-(11):\citet{2000AA...357...608}, (2), (3):\citet{1995AA...297..764}}, (4):\citet{2016aap...A2..23}
\end{deluxetable*}

Solar flares are explosive phenomena wherein magnetic energy stored around sunspots is suddenly released through magnetic reconnection \citep[e.g.,][]{2011shiba...8..6}. Generally, a solar flare releases about 
$10^{26}-10^{32} \: \mathrm{erg}$. Emission in a wide range of wavelengths from radio waves to X-rays occurs during a flare.
Generally, prominence eruptions on the Sun, which are associated with flares \citep[e.g.,][]{2019apj...880..84}, are observed as $\mathrm{H\alpha}$ emission and $\mathrm{H\alpha}$ absorption when they erupt outside a limb and on a disk, respectively \citep{2014solar...11..1, arxiv.2210.02819}.
Prominence or filament eruptions can lead to coronal mass ejections (CMEs) when the prominence velocity is sufficiently large \citep[e.g.,][]{2003apj...586..562, 2011shiba...8..6}.

Flares are widely observed both on the Sun and stars. In the case of stars, “superflares,” which release 10 times larger energies than the largest solar flares, have been confirmed \citep[e.g.,][]{2012nat...485..478}. Spectroscopic observations of stellar flares sometimes show “blue shifts” (or “blue asymmetries”) wherein chromospheric lines during flares are not symmetric, but are enhanced only at shorter wavelengths \citep[e.g.,][]{1990aa...238..249, 1994aaj...285..489, 2004aaj...420..1079, 2008aaj...487..293, 2011aaj...534..A133, 2016aa...590..a11, 2018pasj...70..62, Vida_2019, 2020aa...637..a13, 2021PASJ...73..44-65}. 
An excess component with a shorter wavelength than the rest line center is observed, indicating that the source is flying toward us, considering the Doppler effect. 
Therefore, blue shifts might suggest that an upward-moving plasma exists, such as prominence eruptions \citep{arxiv.2210.02819} or a chromospheric temperature (cool) upflow associated with chromospheric evaporation \citep{10.1093/pasj/psy047}.
One technique to determine whether a blue-shifted excess component originates from cool upflows or prominence eruptions is to estimate it based on its Doppler velocity.
Typically, cool upflows have a velocity of $\sim 100 \: \mathrm{km \: s^{-1}}$ in solar flares \citep[e.g.,][]{2015aa...578..a72}. Therefore, we can expect that blue shifts that significantly exceed this velocity originate from prominence eruptions.

Most of the blue shifts discovered thus far have been relatively slow ($< 500 \: \mathrm{km \: s^{-1}}$). Almost no solid evidence exists that they originated from prominence eruptions and these eruptions triggered a CME.
\citet{2022nature...6..241} first discovered solid evidence of a stellar filament eruption by detecting the blue-shifted absorption component of $\mathrm{H\alpha}$.
By using the length scale and the velocity of the ejected plasma, \citet{2022nature...6..241} confirmed that the erupted filament very likely developed into CMEs.
Further, few cases of blue shifts in flares larger than $10^{35} \: \mathrm{erg}$ emerge.
Generally, a positive correlation exists between the energy of solar flares and those of associated prominence eruptions \citep[e.g.,][]{Aarnio_2011, 2016apjl...833..L8}. Previous blue shift examples found that events on stars generally come as an extension of this correlation \citep[e.g.,][]{2019apj...877..105}.
Therefore, if we can detect a blue shift during a particularly large-scale superflare ($> 10^{35} \: \mathrm{erg}$), its kinetic energy is likely to be excessively large. This becomes reliable evidence of a stellar CME.
Moreover, detection of these events can confirm whether the relation between flare energy and the size of the plasma eruption could be extended to this region.

\epsscale{1.16}
\begin{figure*}[t]
 \begin{center}
  \plotone{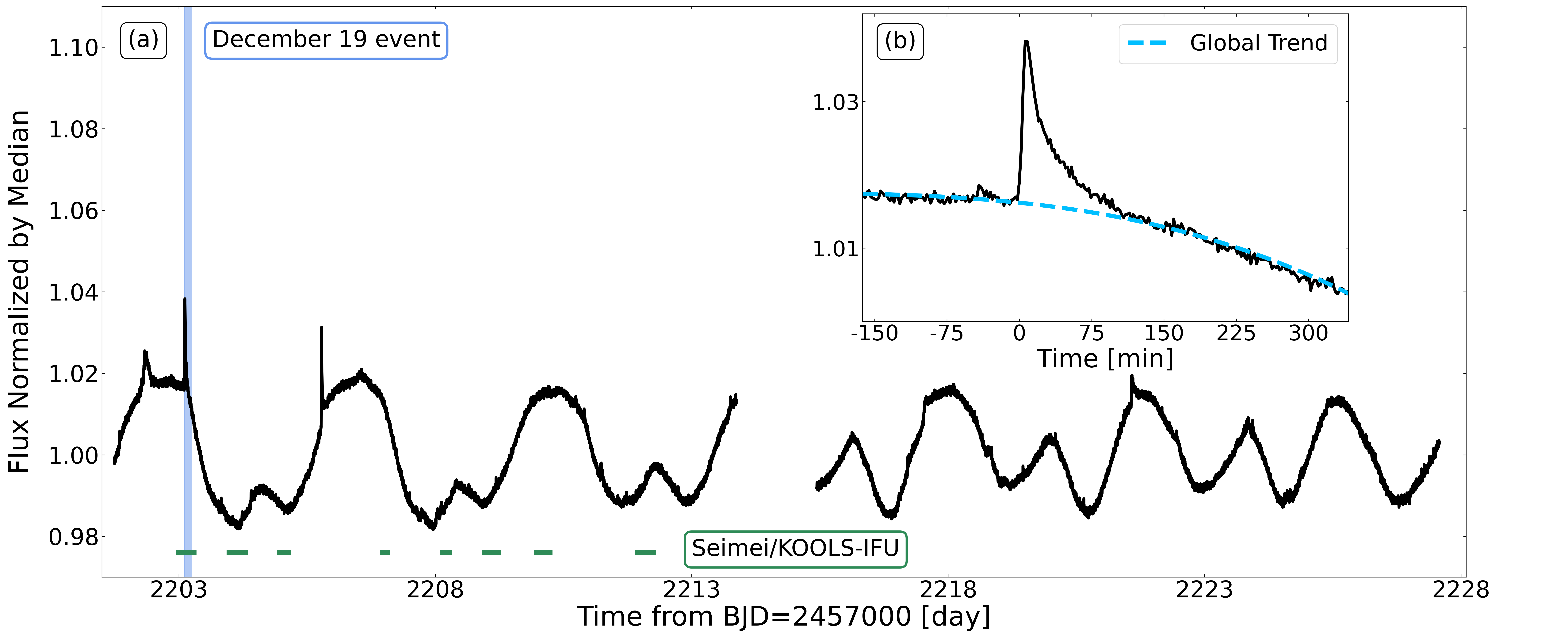}
  \caption{White-light light curves of V1355 Orionis observed with \textit{TESS}. (a) Long-term light curve of V1355 Orionis for BJD=2459201.7-2459227.5. The vertical axis represents the flux normalized by the median value. 
  The vertical light blue line and the horizontal green bars indicate the event discussed in this paper and the period of monitoring observations by Seimei telescope/KOOLS-IFU, respectively.
  (b) Enlarged light curve around December 19/BJD=2459203.11297. The horizontal axis represents the time (unit of minutes) from BJD=2459203.11297. The skyblue dashed line shows the global trend of the stellar rotational modulation fitted for $-500--10 \: \mathrm{min}$ and $150-380 \: \mathrm{min}$ with a polynomial equation.
  \label{fig:tess-long}}
 \end{center}
\end{figure*}

RS CVn-type stars are magnetically active and have been observed to produce large superflares \citep[and references therein]{2016pasj...68..90}.
Many previous studies of flares on RS CVn stars have been conducted in X-ray, whereas only few studies have been conducted with optical spectroscopic observations.
Particulaly, large-scale superflares (i.e., $> 10^{35} \: \mathrm{erg}$) frequently occur on RS CVn stars.
In recent years, stellar CMEs have attracted much attention for their impacts on mass/angular-momentum loss and the environment of surrounding exoplanets. 
If existing correlations between prominence eruptions and flare energies hold for particularly large superflares ($> 10^{35} \: \mathrm{erg}$), prominence eruptions that accompany such flares will lead to particularly large-scale CMEs. Consequently, they will have particularly large effects on stellar evolution and exoplanets \citep{2015apj...809..79, 2020airapetian}.
Therefore, optical spectroscopic observations of RS CVn-type stars are important because they enable the detection of particularly large-scale superflares, which are quite infrequent to be observed on normal main-sequence stars, with only a short period of monitored observations.

In this study, optical spectroscopic observations by the 3.8m  Seimei Telescope \citep{2020PASJ...72..48} and photometric observations by \textit{Transiting Exoplanet Survey Satellite} \citep[\textit{TESS};][]{2015...1..014003} were simultaneously performed to V1355 Orionis, a RS CVn-type star.  
The observations and analysis (Section \ref{sec:Obsevations}), results (Section \ref{sec:Results}), and discussion (Section \ref{sec:Discussion}) on the details of the superflare and the associated blue shift obtained through the simultaneous observations are reported in this paper.

\section{Observations and Analyses} \label{sec:Obsevations}
\subsection{Target star : V1355 Orionis} \label{sec:V1355}
V1355 Orionis (=HD291095) is an RS CVn type binary system
discovered through the ROSAT WFC all-sky X-ray survey \citep{1993MNRAS...260..77, 1995MNRAS...274..1165}.
V1355 Orionis has been investigated by \citet{1995AA...297..764}, 
\citet{1998MNRAS....295...257} and \citet{2000AA...357...608}.
These studies have shown that this binary comprises K0-2IV and G1V stars.
Table \ref{tab:V1355} summarizes the basic physical parameters of the K-type subgiant star.
\citet{2000AA...357...608} reported a large flare on V1355 Orionis in April 1998, which showed 70 times larger equivalent width of $\mathrm{H \alpha}$ than the pre-flare. 

\begin{figure*}[t]
 \begin{center}
  \plotone{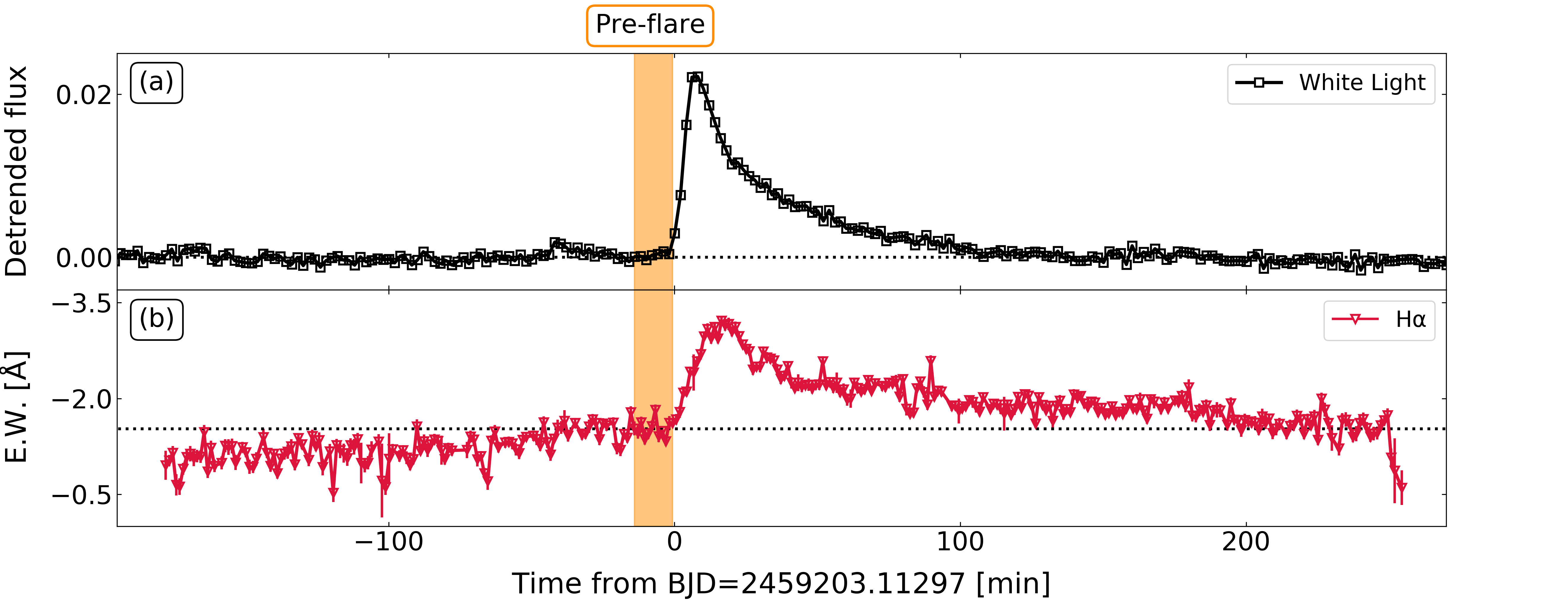}
  \caption{Light curves of V1355 Orionis during the December 19 event. (a) Detrended light curve of the \textit{TESS} white light with background removed. This corresponds to the \textit{TESS} normalized flux (black line in Figure \ref{fig:tess-long} (b)) subtracted by the rotational modulation (skyblue line in Figure \ref{fig:tess-long}(b)). The black dotted line represents the zero level. The orange range indicates the time period determined as the ``pre-flare" in our spectral analysis. (b) Light curve of the $\mathrm{H \alpha}$ for the same time period as in panel (a). The vertical axis shows the equivalent width of $\mathrm{H \alpha}$. The black dotted line represents the level of background, calculated as the average of the equivalent width of the pre-flare.
  Note that in this light curve, the equivalent width is negative for the emission line flux.
  \label{fig:light-curves}}
 \end{center}
\end{figure*}

\begin{figure*}[t]
 \begin{center}
  \plotone{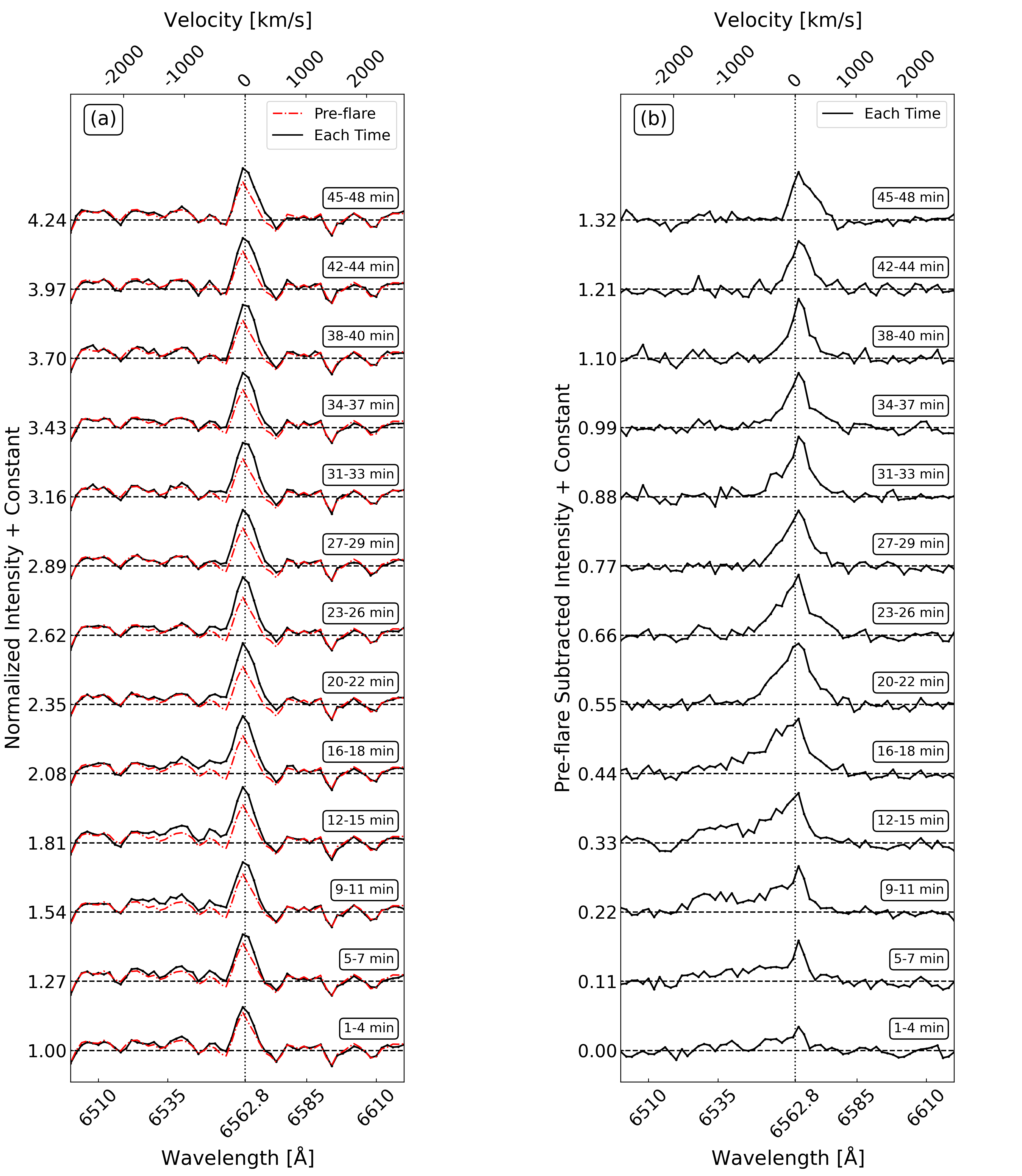}
  \caption{Time variation of the $\mathrm{H \alpha}$ spectrum in the early stages of the flare. (a)
Line profile of $\mathrm{H \alpha}$ emission. 
Each spectrum is a composite of three ($= 3 \: \mathrm{min}$) spectra.
The bottom and top axes are the wavelength and Doppler velocity from the line center, respectively. 
The intensity of each spectrum is normalized by continuum. The vertical dotted and horizontal dashed lines represent the line center of $\mathrm{H \alpha}$ and the continuum level for each spectrum, respectively. 
The spectrum at each time is vertically shifted, and the time shown in the upper right corner of each spectrum 
represents the time from the flare start.
The red dash-dot lines represent the spectra at the pre-flare ($-15$ min to  $0$ min).
(b) Pre-flare subtracted spectrum displayed in the same manner as in panel (a). 
Horizontal dashed line represents the zero level for each spectrum.
The black minus red in panel (a) denotes the spectrum at each time.
 \label{fig:lined-spectrum}}
 \end{center}
\end{figure*}

\subsection{Simultaneous observations}
\subsubsection{Photometric observation : TESS}
\textit{Transiting Exoplanet Survey Satellite} (\textit{TESS})
observed V1355 Orionis in Sector 34 for 27 days with the 2 min time
cadence. Figure \ref{fig:tess-long} (a) shows the \textit{TESS} light curve for this period (BJD=2459201.7-2459227.5).
Figure \ref{fig:tess-long} (b) shows the enlarged \textit{TESS} light curve around the flare described in this paper. The quiescent radiation component during the flare is estimated by fitting a polynomial to the light curve for $-500 - -10 \: \mathrm{min}$ and $+150 - +380 \: \mathrm{min}$ from the flare start (see the skyblue dashed line in Figure \ref{fig:tess-long} (b)).
Figure \ref{fig:light-curves} (a) shows the light curve for the flare component, created by subtracting the quiescent component from the light curve of \textit{TESS} during the flare.
We calculated the bolometric energy of the flare from this detrended light curve following the method of \citet{2013ApJS...209..5}. 
See Section \ref{sec:Results-white} for more details about caluculating the flare energy.
\subsubsection{Spectroscopic observation : the Seimei telescope}
We observed V1355 Orionis using KOOLS-IFU (Kyoto Okayama Optical Low-dispersion Spectrograph with optical-fiber Integral Field Unit; \citet{2019PASJ...71..102}), the spectrograph onboard the 3.8m Seimei telescope \citep{2020PASJ...72..48}.
KOOLS-IFU covers 5800-8000  $\mathrm{\AA}$ with wavelength resolution $R  \: (= \lambda / \Delta \lambda)$ of $\sim2000$. Therefore, it can capture the $\mathrm{H \alpha}$ line.

Observations were made over eight nights in conjunction with Sector 34 of \textit{TESS}. The flare investigated in this study occurred on the first day of our observations. The detailed observation periods are indicated by the horizontal green bars in Figure \ref{fig:tess-long} (a).
The exposure time was set to 60 s on all observation dates.
The signal-to-noise ratio (S/N) at the continuum around $\mathrm{H \alpha}$ line was more than $10$.

Data processing was conducted in the same manner as that of \citet{2020pasj...72..68, 2022nature...6..241, 2022apjl...926..L5}
with \texttt{IRAF} package, \texttt{Pyraf} software, and the data reduction packages developed by \citet{2019PASJ...71..102}. Using the $\mathrm{H \alpha}$ line profile at each time frame, we investigated the time variation of the equivalent width during the flare. Figure \ref{fig:light-curves} (b) shows the light curve of the equivalent width of $\mathrm{H \alpha}$. Further, the equivalent width was calculated, after normalizing $\mathrm{H \alpha}$ emission by the nearby continuum level, integrating it for $6518-6582 \: \mathrm{\AA}$. The reason why the range of integration is asymmetric to the line center of $\mathrm{H \alpha} \: (=6562.8 \: \mathrm{\AA})$ is due to the blue-shifted excess component. See Section \ref{sec:Results-Ha} for more information about the blue-shifted excess component.

\section{Results} \label{sec:Results}
\subsection{White light flare energy} \label{sec:Results-white}
As shown in Figure \ref{fig:light-curves} (a), the white light flare lasted about $110 \: \mathrm{min}$.
By setting the flux of Figure \ref{fig:light-curves} (a) to $C'_{\mathrm{flare}}$ ($=$ luminosity of flare$/$luminosity of star) in the equation (5) of \citet{2013ApJS...209..5}, we estimated the flare area $A_{\mathrm{flare}}$ as follows,
\begin{equation} 
A_{\mathrm{flare}}(t) = \frac{\pi C'_{\mathrm{flare}}(t) \sum_{i=1, 2} \left \{ R_{i}^2 \int R_{\lambda} B_{\lambda} (T_{i}) d \lambda \right \}}{\int R_{\lambda} B_{\lambda} (T_{\mathrm{flare}}) d \lambda } 
\end{equation} 
where $\lambda$ is the wavelength, $B_{\lambda}$ is the Planck function, $R_{\lambda}$ is the \textit{TESS} response function \citep{2015...1..014003}, and $T_{i}$ is the effective temperature of the K- and G-type stars of the binary \citep[$4750 \: \mathrm{K}$ and $5780 \: \mathrm{K}$, respectively;][]{2000AA...357...608}. Further, $T_{\mathrm{flare}}$ is the flare temperature of $10000 \: \mathrm{K}$ \citep{1980apj...239..L27, 1992apsj...78..565}, and $R_{i}$ is the radius of the K- and G-type stars of the binary \citep[$4.1 R_{\odot}$ and $1.0 R_{\odot}$, respectively;][]{2000AA...357...608}.
Assuming that the flare can be approximated by blackbody radiation with a temperature of $T_{\mathrm{flare}} = 10000 \: \mathrm{K}$, flare luminosity $L_{\mathrm{flare}}$ is
\begin{equation}
    L_{\mathrm{flare}}(t) = \sigma_{\mathrm{SB}} T_\mathrm{flare}^4 A_{\mathrm{flare}}(t)
\end{equation}
where $\sigma_{\mathrm{SB}}$ is the Stefan-Boltzmann constant.
Finally, by integrating $L_{\mathrm{flare}}$ with the duration of the white light flare ($\sim 110 \: \mathrm{min}$), the bolometric flare energy $E_{\mathrm{bol}}$ is
\begin{equation} 
    E_{\mathrm{bol}} = \int L_{\mathrm{flare}}(t) \: dt = 7.0 \times 10^{35} \; \mathrm{erg}. 
\end{equation} 
Given that the K-type star is more magnetically active and the bolometric flare energy is very large, we consider this flare to have occurred on the K-type star of the binary.
\begin{figure*}[t]
 \begin{center}
  \plotone{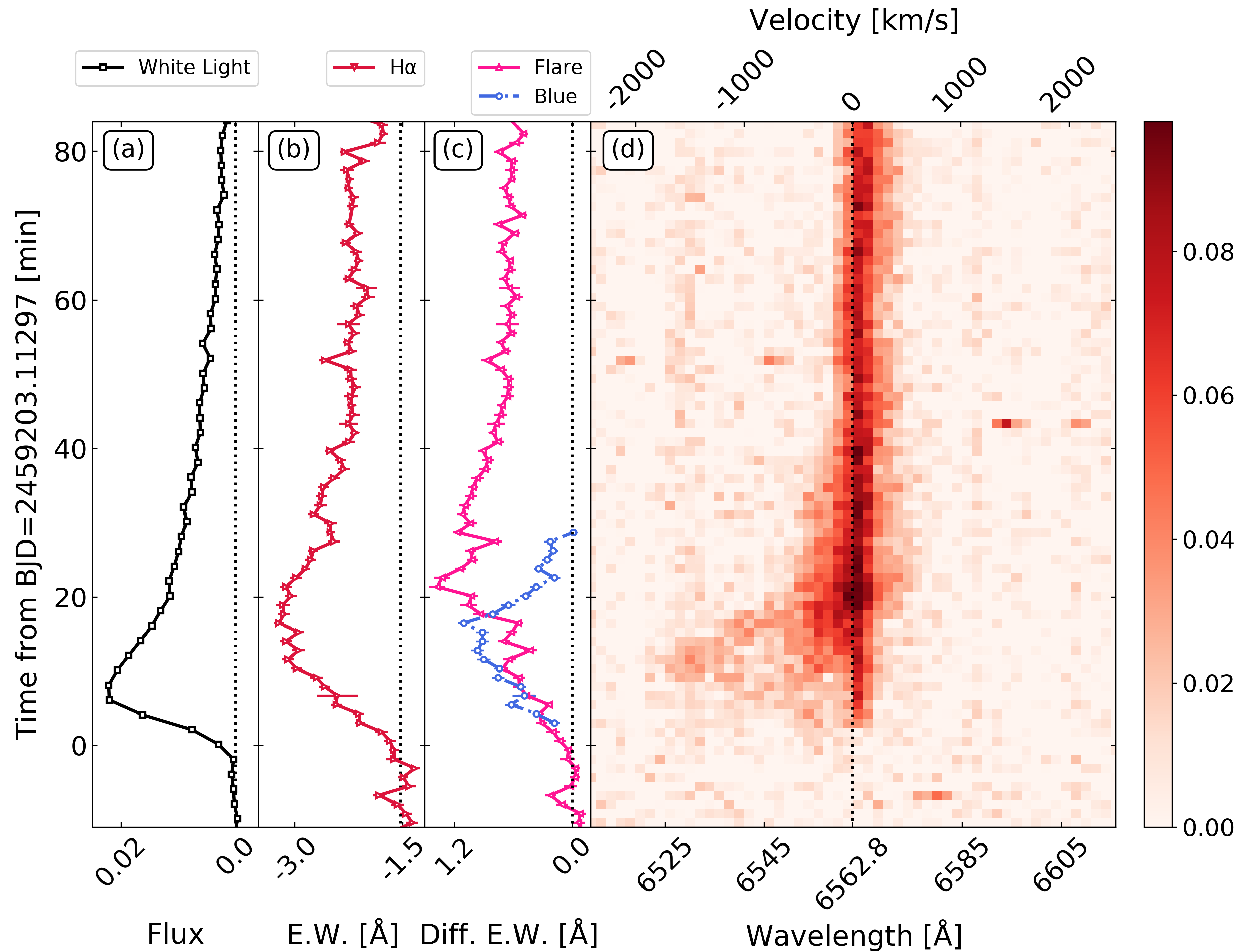}
  \caption{Light curves and spectra during a superflare on V1355 Orions. (a) Enlarged early part of the light curve of Figure \ref{fig:light-curves} (a). The horizontal and vertical axes indicate the detrended flux and time (unit of minutes) from BJD=2459203.11297, respectively. (b) Enlarged $\mathrm{H \alpha}$ light curve of Figure \ref{fig:light-curves} (b) for the same period as in panel (a).
The horizontal and vertical axis represents the equivalent width of $\mathrm{H \alpha}$  and that the same as in (a), respectively. (c)$\mathrm{H \alpha}$ light curve divided into line center and blue-shifted excess components. The pink triangles and blue circles indicate the equivalent widths of the flare and blue-shifted excess components, respectively. 
 The equivalent width
  of the flare component was calculated by integrating the differential spectrum from the pre-flare level
 $\pm 17 \: \mathrm{\AA}$ 
 from the $\mathrm{H \alpha}$ line center while blue-shifted excess component was not present. For the time period when the emission line was blue-shifted,  the equivalent width was calculated by integrating the Voigt function that fits the spectrum at longer wavelengths than the peak (e.g., the black dotted line in Figure \ref{fig:blue-asymmetry}(a)).
The equivalent width of the blue-shifted excess component was calculated by integrating the residuals between the
spectra and the fitted Voigt functions (e.g., the blue line in Figure \ref{fig:blue-asymmetry}(b)).
See Section \ref{sec:estimation} for more information about fitting. (d) Time variation of the differential spectrum from the pre-flare level at the same time as in (a), (b), and (c). The bottom and top abscissas are the wavelength and Doppler velocity from the line center, respectively.
 \label{fig:color-map}}
 \end{center}
\end{figure*}

\epsscale{0.8}
\begin{figure*}[t]
 \begin{center}
 \plotone{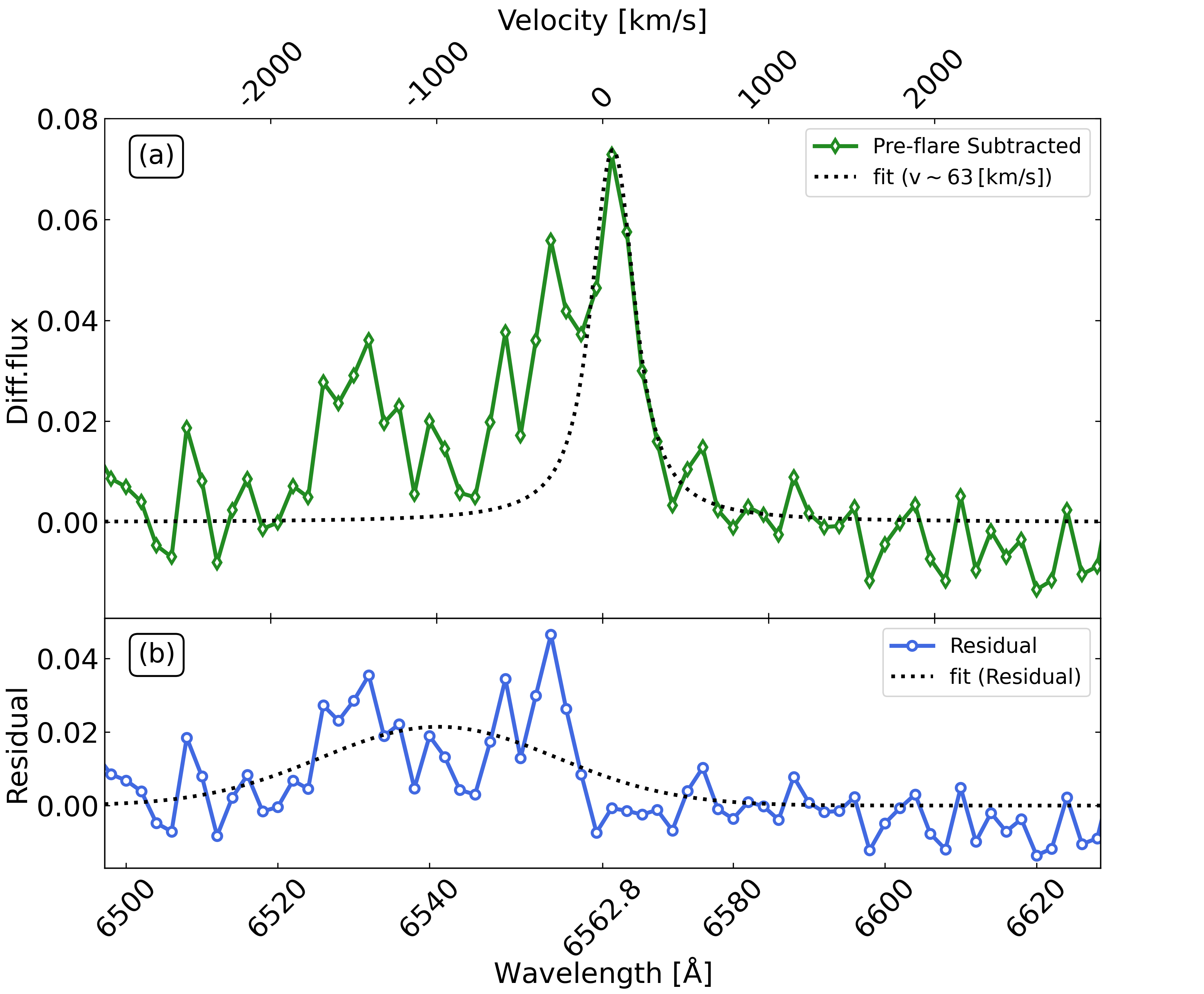}
 \plotone{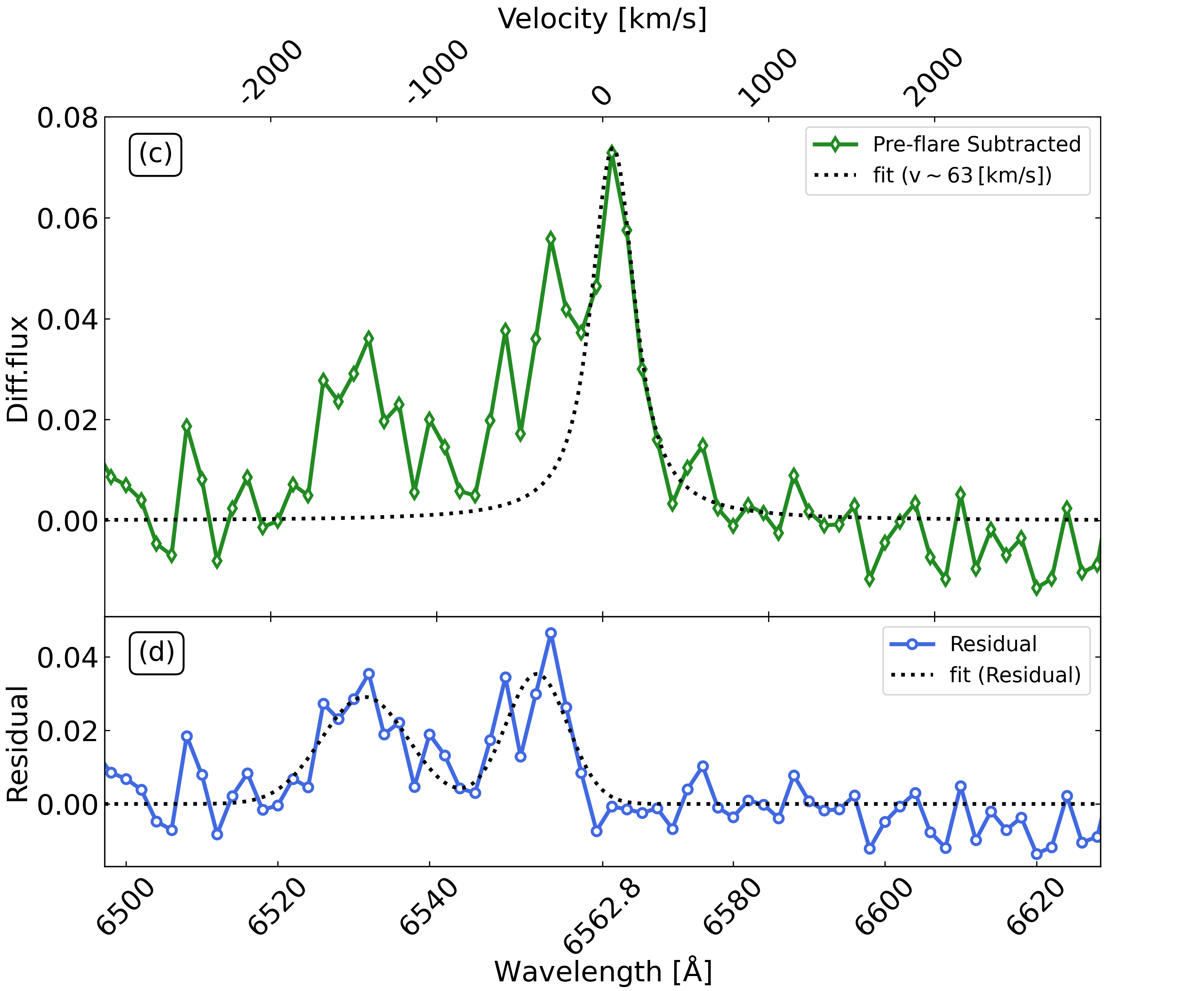}
  \caption{Pre-flare subtracted spectrum and its residual difference from Voigt fitting for the time period when the blue-shifted excess component appeared most prominently composed of two components. (a), (c) Pre-flare subtracted spectrum for BJD = 2459203.12016776.
The black dotted line represents the Voigt function fitted only for the longer wavelength side of the line center.
Note that the center of the Voigt function is set to the value of the radial velocity of V1355 Orionis calculated from the rotational phase. (b), (d) Residual between the observed spectrum and fitting.
These spectra correspond to the green line minus black dotted line in (a), (c).
The black dotted line represents the Gaussian with the residual fitted, shown  with (b) one- and (d)two-component.
  \label{fig:blue-asymmetry}}
 \end{center}
\end{figure*}

\epsscale{1.16}
\begin{figure*}[t]
 \begin{center}
  \plotone{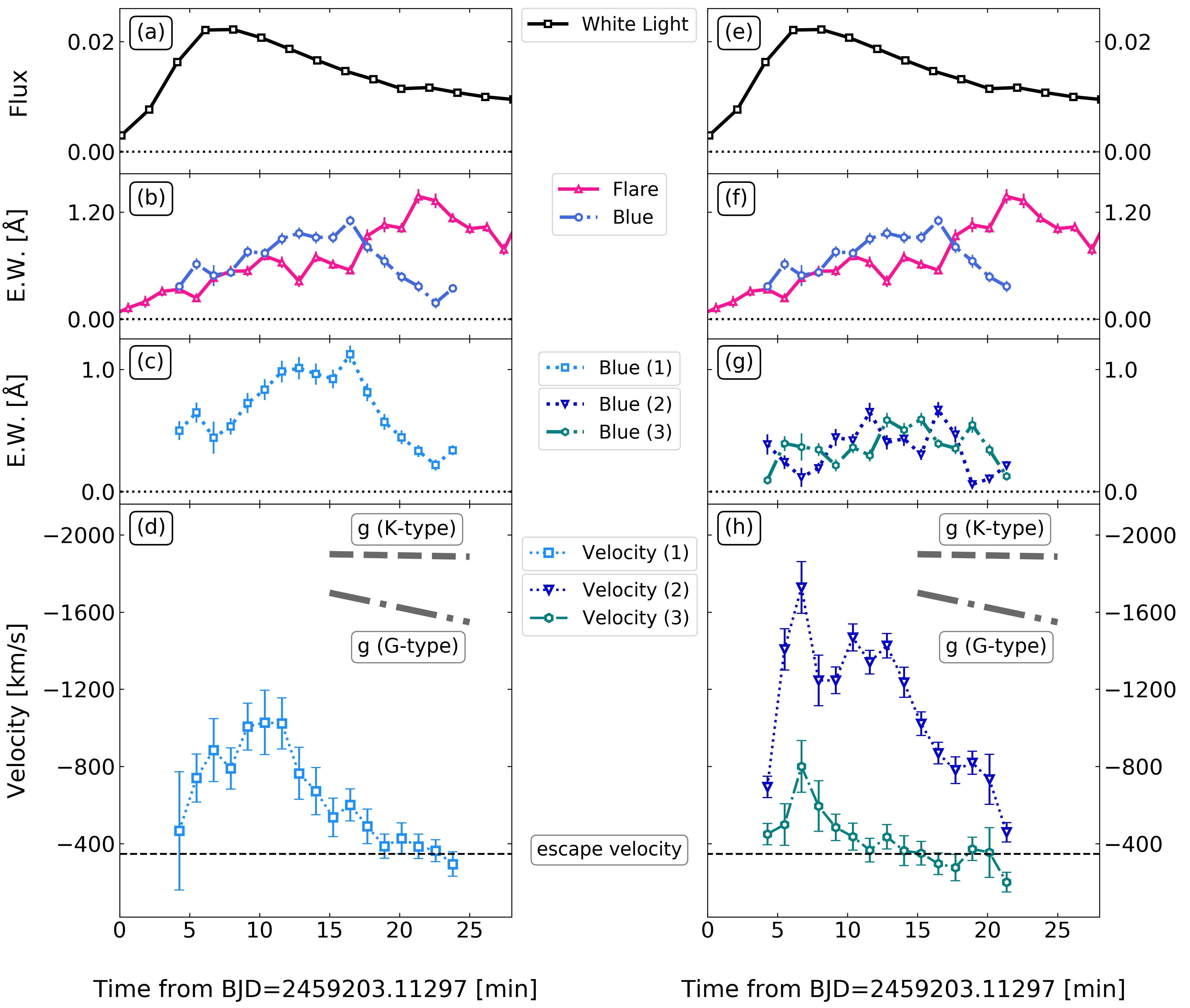}
  \caption{Light curves and time variation of the velocity of the blue-shifted excess component. (a),(e) Light curves of white light observed with \textit{TESS}; this is an enlarged version of the light curve shown in Figure \ref{fig:light-curves} (a) only for the time period with the blue-shifted excess component visible. (b),(f) Light curves of the equivalent width of $\mathrm{H \alpha}$ observed with KOOLS-IFU on the Seimei telescope at the same time as (a), (e). The pink triangles and blue circles indicate the equivalent widths of the flare and blue-shifted excess components, respectively. 
(c),(g) Light curves of the equivalent widths of the components comprising the residuals between the Voigt function and pre-flare subtracted spectrum.
These values were calculated by integrating Gaussian functions fitted with residuals.
Light blue squares denotes the equivalent width when fitting the residuals with one component.(Blue (1))
Medium blue inverted triangles and teal hexagons represent the equivalent widths of the faster and slower components, respectively, when the residual is fitted by two components (Blue (2)/(3)).
(d),(h) Time variation of the velocity of the blue-shifted excess components.
Marks are set as in panel (c), (g).
The error bars contain two elements: fitting error and the variation of the radial velocity of this star.
The slopes of the gray lines represent the gravitational acceleration at the surface of each of the binary stars. The black dashed lines show the escape velocity of the K-type 
star of V1355 Orionis.
 \label{fig:blue-velocity}}
 \end{center}
\end{figure*}

\epsscale{0.9}

\begin{figure*}[t]
 \begin{center}
  \plotone{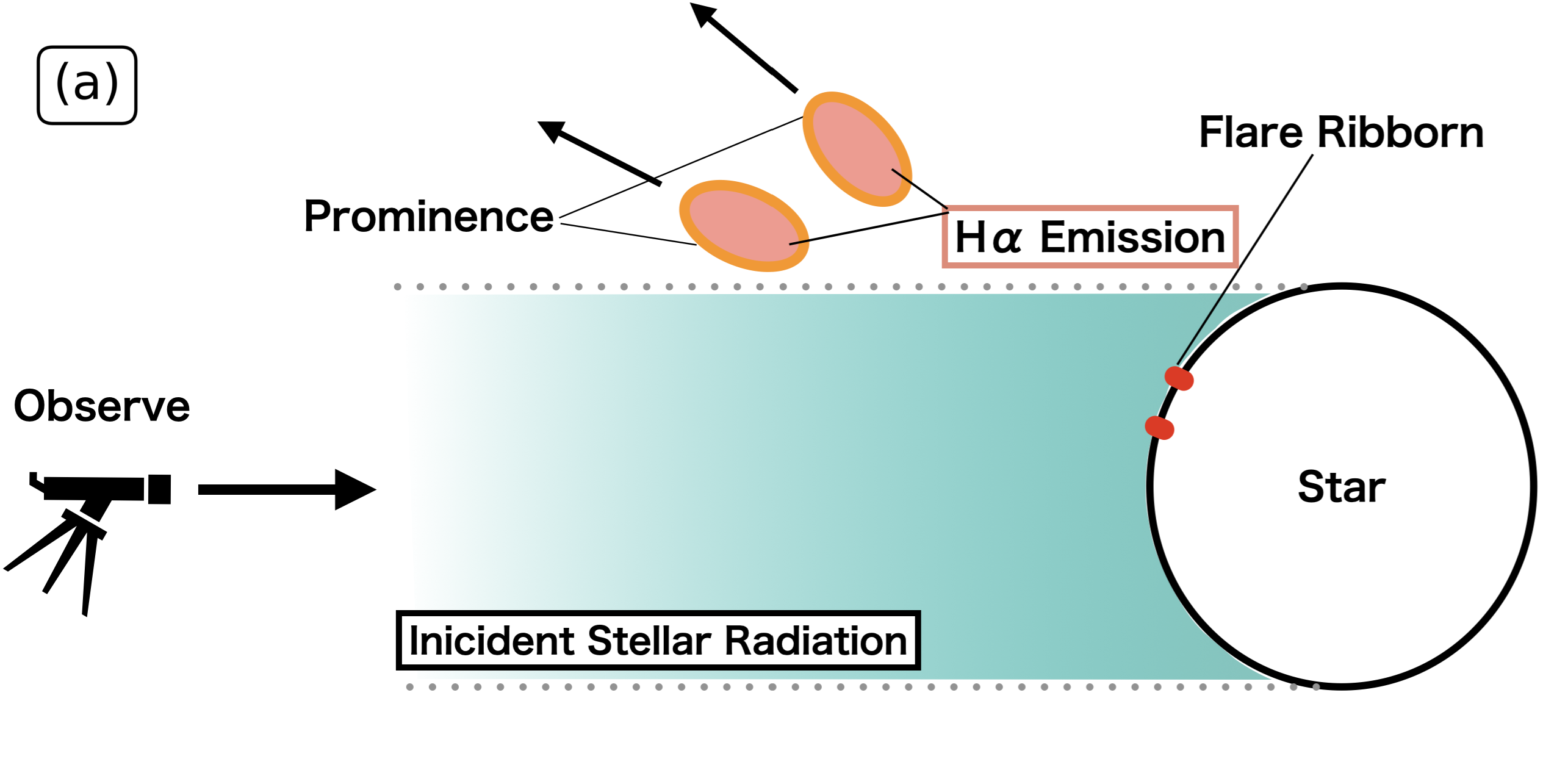}
  \plotone{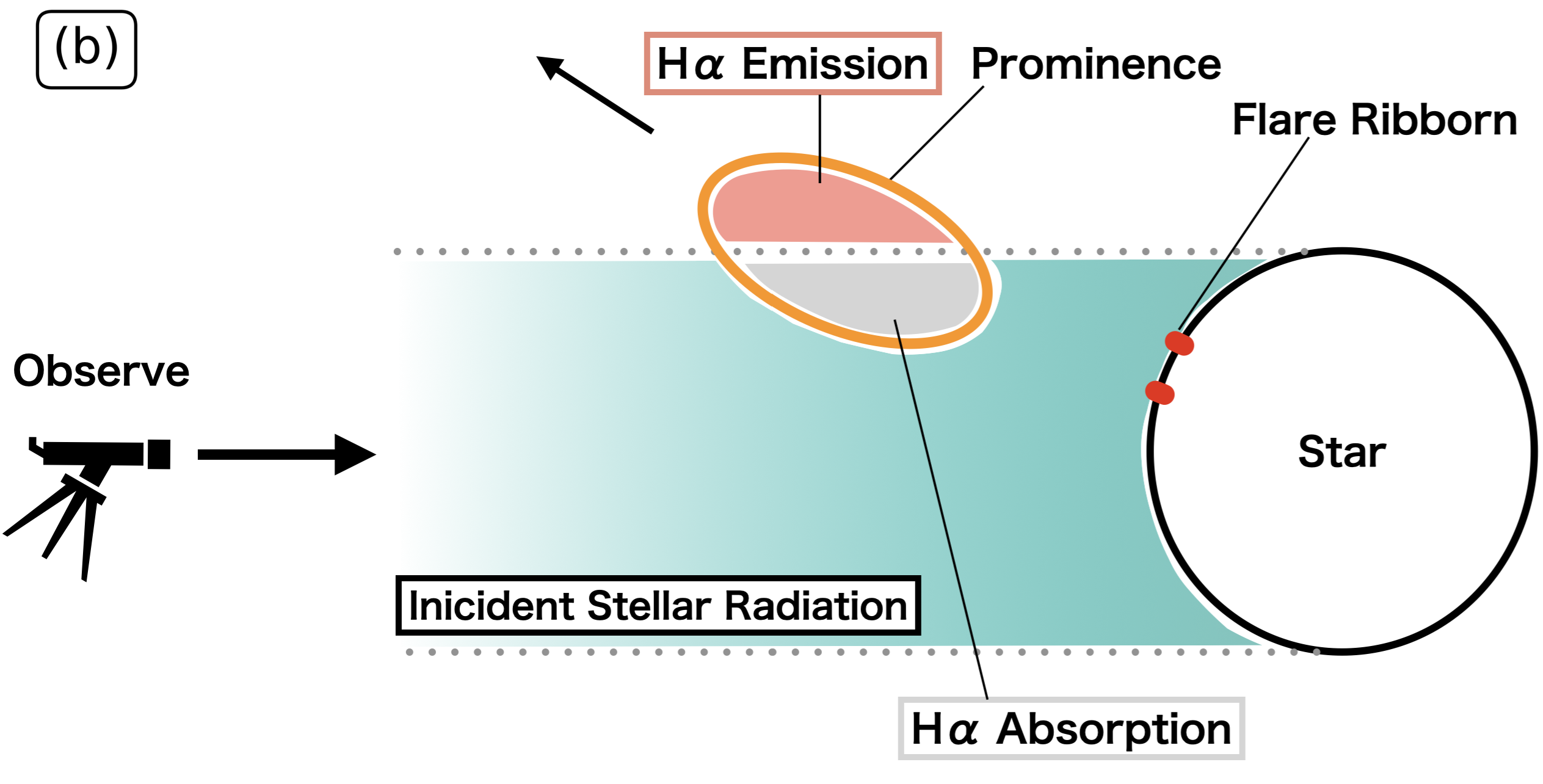}
  \caption{Schematic diagram that represents the interpretation of the fact that the blue-shifted excess component appears to be composed of two components.
  (a) When two prominences are visible in the $\mathrm{H \alpha}$ emission line.
  (b) When parts of the prominence are visible in the $\mathrm{H \alpha}$ emission line and parts are visible in absorption. 
 \label{fig:kaisyaku}}
 \end{center}
\end{figure*}

\subsection{H$\alpha$ line profile during the flare} \label{sec:Results-Ha}
Our spectroscopic observation revealed that a remarkable blue-shifted excess component exists in the $\mathrm{H \alpha}$ emission line for $\sim 30$ min after the start of the flare. Figure \ref{fig:lined-spectrum} (a) shows the spectra during the time when the blue-shifted excess component was observed during the flare.
We estimated the spectrum of the quiescent component, denoted by red spectrum in Figure \ref{fig:lined-spectrum} (a),  by combining the spectra from 15 min before the flare start (“pre-flare”). 
Figure \ref{fig:lined-spectrum} (b) shows the spectrum of the flare emission component, which were obtained by obtaining the difference between the spectrum at each time and the pre-flare spectrum.
A particularly large blue-shifted excess component was observed in the pre-flare subtracted spectrum, especially $5-18 \: \mathrm{min}$, extending over $-1000 \:  \mathrm{km \: s^{-1}}$. Figure \ref{fig:color-map} (d) presents a color map showing the time variation of the blue-shifted excess component in the pre-flare subtracted spectrum.
The blue-shifted excess component was fast over the time period when white light and $\mathrm{H \alpha}$ reach their peak.

While the spectrum before the difference did not do so (Figure \ref{fig:lined-spectrum} (a)), the $\mathrm{H \alpha}$ peak of the pre-flare subtracted spectrum was slightly red-shifted ($\sim +50 \: \mathrm{km \: s^{-1}}$; Figure \ref{fig:lined-spectrum} (b) and \ref{fig:color-map} (d)).
All spectra we present in this paper are not corrected for the radial velocity.
According to \citet{2000AA...357...608}, the radial velocity of V1355 Orionis varies in the range of about $+0-+70 \: \mathrm{km \: s^{-1}}$.
That is, the velocity of the red shift is within the range of the radial velocity variability.
The red shift might be caused by the downward chromospheric condensation \citep{1984/10.1007/BF00156656} or the post flare-loop \citep{2019...AA..624..A96}.

We divided the $\mathrm{H \alpha}$ emission line into flare and blue-shifted excess components and then calculated their equivalent widths. Figure \ref{fig:color-map} (c) shows their light curves.
The emission lines were separated into flare and blue-shifted excess components by fitting only the long wavelength side of the lines with the Voigt function. See Section \ref{sec:estimation} for details about the fitting.
From the light curve separated into flare and blue-shifted excess components, we calculated the energy of the flare emitted in $\mathrm{H \alpha}$ ($E_{\mathrm{H \alpha}}$). 
Using the \textit{R-}band magnitude \citep[$m_{R}$;][]{1995AA...297..764}, $R$-band Vega flux zero point per unit wavelength \citep[$f_{\lambda} = 217.7 \times 10^{-11} \: \mathrm{ergs \: cm^{-2} \: sec^{-1} \: \AA^{-1}}$;][]{1998aap...333..231}, and the distance between the Earth and V1355 Orionis \citep[$d$;][]{2016aap...A2..23}, the equivalent width of the flare ($EW_{\mathrm{flare}}$) can be converted to luminosity $L_{\mathrm{H \alpha}}$, 
\begin{equation}
\label{eq:L_EW}
    L_{\mathrm{H \alpha}} (t) = 4 \pi d^2 f_{\lambda} 10^{-0.4m_{R}} \times EW_{\mathrm{flare}} (t).
\end{equation}
Table \ref{tab:V1355} lists the values of $d$ and $m_{R}$.
By integrating $L_{\mathrm{H \alpha}}$ with the duration of the $\mathrm{H \alpha}$ flare ($\sim 210 \: \mathrm{min}$), 
\begin{equation}
    E_{\mathrm{H \alpha}} = \int L_{\mathrm{H \alpha}} (t) \: dt = 1.1 \times 10^{34} \; \mathrm{erg}
\end{equation}
is obtained.
\section{Discussion} \label{sec:Discussion}
\subsection{Estimation of the prominence parameters}\label{sec:estimation}
As discussed in Section \ref{sec:Results-Ha}, a clear blue-shifted excess component in the $\mathrm{H \alpha}$ emission line was identified during the flare. 
This blue-shifted excess component must reflect a prominence eruption since its velocity extends to more than $1000 \: \mathrm{km \: s^{-1}}$.
A cool upflow \citep{10.1093/pasj/psy047} cannot explain such a rapid blue shift.
Therefore, we estimated the velocity and mass of the prominence that appears as the blue-shifted excess component.
\subsubsection{Velocity}
We estimated the velocity of the prominence using a method similar to that used by \citet{2021PASJ...73..44-65}.
As shown in Figure \ref{fig:blue-asymmetry} (a) and (c), we first fit the pre-flare subtracted spectrum at each time with the Voigt function only at wavelengths longer than the line center. Then, we calculated the residuals between the Voigt function and the pre-flare subtracted spectrum. The residual is shown by the blue line in Figure \ref{fig:blue-asymmetry} (b) and (d). Finally, the residual was fitted with the Gaussian. The wavelength of the Gaussian peak was converted to the Doppler velocity to evaluate the prominence velocity.

Spectra exist over multiple time periods when the residual appeared to have two peaks, as shown in Figure \ref{fig:blue-asymmetry} (b) and (d).
Therefore, when fitting the residuals, we performed two types of fits: one- and two-component Gaussian fits (Figure \ref{fig:blue-asymmetry} (b) and (d), respectively).
See Section \ref{sec:two-components} for the details of and interpretation on why the blue-shifted excess component appearing to have two peaks.

The time variation of the velocity of the prominence examined in this manner is shown in Figure \ref{fig:blue-velocity} (d) and (h).
In the case of a one-component fit, the velocity of the prominence reached $-990 \pm 130 \: \mathrm{km \: s^{-1}}$ at the peak.
In the case of a two-component fit, the velocity of the prominence reached $-1690 \pm 100 \: \mathrm{km \: s^{-1}}$ and $-760 \pm 90 \: \mathrm{km \: s^{-1}}$ at the peak.
In the case of both fits, the prominence velocity was almost always much faster than the escape velocity ($-347 \: \mathrm{km \: s^{-1}}$) at the surface of the K-type star of V1355 Orionis.
Therefore, the prominence that erupted with this flare would certainly have flown outward from the star and developed into a CME.

Moreover, in both cases, the velocity of the blue-shifted excess component decelerated more rapidly than the gravitational acceleration of the K-type star after reaching its peak (see gray lines in Figure \ref{fig:blue-velocity} (d) and (h)).
The following are the two possible interpretations.
The first interpretation is that what we observe in Figure \ref{fig:blue-velocity} (d) and (h) is not the actual deceleration.
The fast part of the erupted prominence becomes invisible rapidly, and the slow part becomes gradually dominant. This is observed in the Sun-as-a-star analysis of a filament eruption analyzed by \citet[][see the supplementary information therein]{2022nature...6..241}.
The second interpretation is that the magnetic field applies force to the prominence in addition to gravity.
Simulations conducted by \citet{Alvarado-Gomez_2018} have shown that the magnetic field of an active star could contribute significantly to the slowing of prominences.
\subsubsection{Mass}
\label{sec:Mass}
We estimated the upper and lower limit prominence mass from the equivalent width of the blue-shifted excess component. In the method used by \citet{2021PASJ...73..44-65}, the upper limit of prominence mass is proportional to the 1.5 power of the area of the region emitting $\mathrm{H \alpha}$. Therefore, for a prominence eruption as large in scale as this case, the method can significantly overestimate the upper limit of prominence mass because the prominence shape would be far from cubic like solar prominences. So, we improved the method used by \citet{2021PASJ...73..44-65}. 

As shown in Figure \ref{fig:color-map} (c), the maximum equivalent width of the blue-shifted excess component is $\sim1 \: \mathrm{\AA}$. 
Converting the equivalent width to luminosity using equation (\ref{eq:L_EW}), the luminosity of the blue-shifted excess component $L_{\mathrm{blue}}$ is obtained as
\begin{equation}
\label{eq:L_blue}
    L_{\mathrm{blue}} \sim 1 \times 10^{30} \; \mathrm{erg \: s^{-1}}.
\end{equation}
We assume that the NLTE model of the solar prominence \citep{1994aaj...292..656} can be adapted to the present case, and further assume that the optical thickness of $\mathrm{H \alpha}$ line center $\tau_{\mathrm{p}}$ is $0.1 - 100$. 
\begin{itemize}
\item $\tau_{\mathrm{p}} \sim 0.1$: \\
The $\mathrm{H\alpha}$ flux of the prominence per unit time, unit area, and unit solid angle $F_{\mathrm{H\alpha}}$ is 
\begin{equation}
\label{eq:F_value} 
    F_{\mathrm{H\alpha}} \sim 10^{4} \; \mathrm{erg \: s^{-1} \: cm^{-2} \: sr^{-1}} 
\end{equation}
\citep[see Figure 5 in ][]{1994aaj...292..656}.
As the integral of $F_{\mathrm{H\alpha}}$ over the region emitting the $\mathrm{H\alpha}$ and the solid angle in the direction toward us is $L_{\mathrm{blue}}$, 
\begin{equation}
\label{eq:L_blue2} 
    L_{\mathrm{blue}} = \int \int F_{\mathrm{H\alpha}} \: dA d\Omega = 2 \pi A F_{\mathrm{H\alpha}} 
\end{equation}
where $A$ is the area of the region emitting $\mathrm{H\alpha}$.
From equations (\ref{eq:L_blue})$-$(\ref{eq:L_blue2}), 
\begin{equation}
\label{eq:A_value} 
    A \sim 5 \times 10^{24.5} \; \mathrm{cm^{2}} 
\end{equation}
is obtained. From equation (\ref{eq:F_value}), the emission measure $n_{\mathrm{e}}^{2} D$ of the prominence is 
\begin{equation} 
n_{\mathrm{e}}^{2} D \sim 10^{28} \; \mathrm{cm^{-5}} \label{eq:EM_value} 
\end{equation}
\citep[see Figure 15 in ][]{1994aaj...292..656}
where $D$ and $n_{\mathrm{e}}$ are the geometrical thickness and the electron density of the prominence, respectively. Though \cite{1994aaj...292..656} assumes a range of $D$, $F_{\mathrm{H \alpha}}$ and $n_{\mathrm{e}}^{2} D$ are largely uniquely determined for a value of $\tau_{\mathrm{p}}$ without much influence from the indefiniteness of $D$. On the other hand, we need to assume values of the electron density $n_{\mathrm{e}}$ and the hydrogen density $n_{\mathrm{H}}$.
The typical electron density of solar prominence is
\begin{equation}
n_{\mathrm{e}} \sim 10^{10-11.5} \; \mathrm{cm^{-3}} \label{eq:ne_value}
\end{equation}
\citep{1986Hirayama}.
From equations (\ref{eq:EM_value}) and (\ref{eq:ne_value}), 
\begin{equation}
\label{eq:D_value} 
D \sim 10^{5-8} \; \mathrm{cm}. 
\end{equation}
\citet{Labrosse1994} showed the ratio between the hydrogen density $n_{\mathrm{H}}$ and the electron density $n_{\mathrm{e}}$ of solar prominence is
\begin{equation}
n_{\mathrm{e}} / n_{\mathrm{H}} \sim 0.2-0.9. \label{eq:nenh_ratio}
\end{equation}
From equations (\ref{eq:A_value}), (\ref{eq:ne_value}), (\ref{eq:D_value}) and (\ref{eq:nenh_ratio}), 
the mass of the prominence 
\begin{equation}
M \sim m_{\mathrm{H}} n_{\mathrm{H}} A D
\end{equation}
is 
\begin{equation} 
\label{eq:M_value_01}
9.5 \times 10^{18} \; \mathrm{g} < M < 1.4 \times 10^{20} \; \mathrm{g}
\end{equation} 
where $m_{\mathrm{H}}$ is the mass of hydrogen atom. The error range comes from the assumed range of electron density and degree of ionization.
\item $\tau_{\mathrm{p}} \sim 100$: \\
When the value of $\tau_{\mathrm{p}}$ is $\sim 100$, 
\begin{equation}
\label{eq:F_value_100} 
    F_{\mathrm{H\alpha}} \sim 10^{6} \; \mathrm{erg \: s^{-1} \: cm^{-2} \: sr^{-1}} 
\end{equation}
\citep[see Figure 5 in ][]{1994aaj...292..656}.
Calculated as in case $\tau_{\mathrm{p}} \sim 0.1$, 
\begin{equation}
A \sim 1.6 \times 10^{23} \; \mathrm{cm^{2}}. 
\end{equation}
Assuming equations (\ref{eq:ne_value}) and (\ref{eq:nenh_ratio}) as in case $\tau_{\mathrm{p}} \sim 0.1$, 
\begin{equation}
D \sim 10^{8-11} \; \mathrm{cm} 
\end{equation}
\begin{equation} \label{eq:M_value_100}
9.5 \times 10^{19} \; \mathrm{g} < M < 1.4 \times 10^{21} \; \mathrm{g}.
\end{equation}
\end{itemize} 
Combining the ranges of $M$ in equations (\ref{eq:M_value_01}) and (\ref{eq:M_value_100}),
\begin{equation}
9.5 \times 10^{18} \; \mathrm{g} < M < 1.4 \times 10^{21} \; \mathrm{g}.
\end{equation} 

As shown in equations (\ref{eq:M_value_01}) and (\ref{eq:M_value_100}), the upper and lower limits of the prominence mass varied by only about an order of magnitude when $\tau_{\mathrm{p}}$ varied significantly.
Since we do not know the value of $\tau_{\mathrm{p}}$, we set an extreme value of $\tau_{\mathrm{p}} \sim 100$ as the upper limit in this paper.
The energy of this flare is $\sim 10^{6}$ times that of a typical solar flare ($\sim 10^{30} \: \mathrm{erg}$). The energy of a flare is proportional to the cube of the spatial scale \citep{Shibata_2002}. The typical value of the optical thickness of the solar prominence is $\sim 1$. Simply put, the optical thickness of the prominence is proportional to the geometric thickness of the prominence. Given the spatial scale of the prominence, the upper limit of $\tau_{\mathrm{p}}$ can be roughly considered to be $\sim 1 \times (10^{6})^{1/3} = 100$.
On the other hand, the lower limit of $\tau_{\mathrm{p}}$ was limited by the hemispheric area of the star.
When $\tau_{\mathrm{p}}$ is set to an extremely small value, $A$ is extremely larger than the hemispherical area of the star (see equations (\ref{eq:L_blue}) and (\ref{eq:F_value})). Such a situation is unrealistic. The value of $A$ in equation (\ref{eq:A_value}) corresponds to $\sim 10^{1.5} \pi R^{2}$.
Filling factor effect may also affect the prominence mass estimation \citep{Kucera_1998}. So, a more detailed study of the stellar prominence mass calculation is needed in the future.

\epsscale{1.16}
\begin{figure*}[t]
 \begin{center}
  \plotone{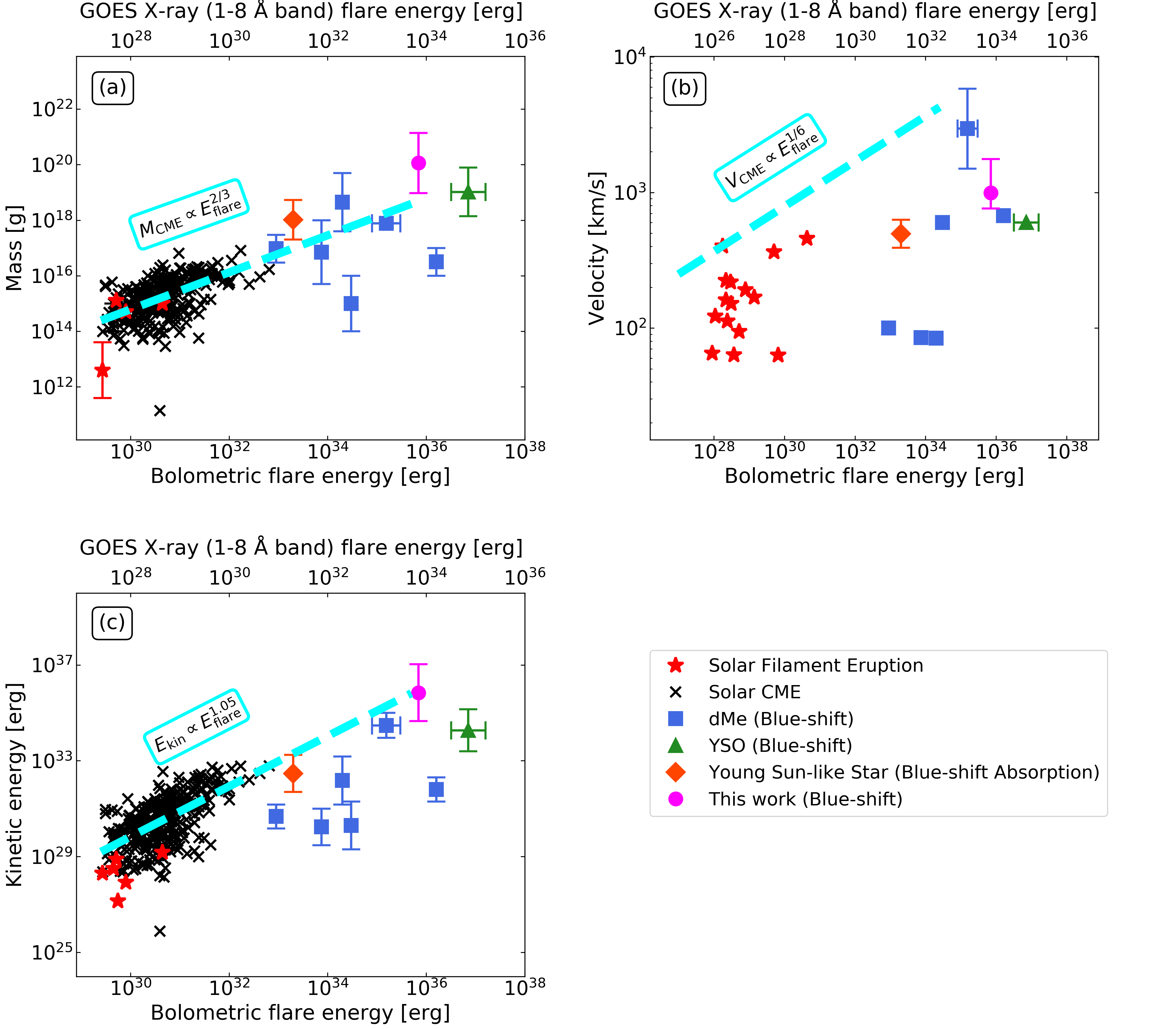}
  \caption{Mass, kinetic energy, and velocity of the prominence eruption on V1355 Orionis compared with the statistical data of solar and stellar prominence eruptions/CMEs. (a) Comparison between mass of CMEs/prominences and flare energy. The top and bottom horizontal axes represent the energy emitted in the GOES wavelength band and bolometric flare energy, respectively. 
  Red stars correspond to filament eruptions on the Sun taken from \citet{2022nature...6..241}.
  Black crosses correspond to CME events on the Sun taken from \citet{2009univ...257..233}. Blue squares indicate stellar mass ejection events on M-dwarfs (dMe). Green triangles indicate stellar mass ejection events on young stellar objects (YSO) and close binary systems (CB). Data of these stellar events were obtained from \citet{2019apj...877..105} and \citet{2021PASJ...73..44-65}. The orange diamond represents the filament eruption event on a young Sun-like star EK Dra reported in \citet{2022nature...6..241}, wherein a blue-shifted absorption component was identified. 
  The pink circle denotes the prominence eruption on V1355 Orionis. 
  The cyan dashed line represents the relation: $M_{\mathrm{CME}} \propto E^{2/3}_{\mathrm{flare}} $, shown by \citet{2016apjl...833..L8} about the Sun, which is fitted to the solar data points used in this study.
  (b) Velocity of CMEs/prominences as a function of flare energy. 
  Red stars represent the filament eruptions on the Sun (obtained from \citet{2019...14..95}). The pink circle denotes the velocity of the prominence eruption on V1355 Orionis obtained by fitting with one-component Gaussian. The lower and upper limits of the velocity of the prominence eruption on V1355 Orionis are the velocity obtained by fitting with two-component Gaussian. The other marks are the same as in (a).
  The scaling law denoted by the cyan dashed line was obtained from \citet{2016apjl...833..L8}. Note that this scaling law is an upper bound on the speed; thus it has been adjusted to pass through the fastest point in the solar data used here.
  (c) Comparison between kinetic energy of CMEs/prominences and flare energy. The horizontal axis and each mark are the same as those in (a). 
  The scaling law denoted by the cyan dashed line was obtained from \citet{2022nature...6..241}, which is also fitted to the solar data points used in this study.
  \label{fig:statistics}}
 \end{center}
\end{figure*}

\subsection{Interpretation of $\mathrm{H \alpha}$ line profile: Two components}\label{sec:two-components}
Sometimes the blue-shifted excess component appeared to have two clear peaks, as shown in Figure \ref{fig:blue-asymmetry} (b) and (d), whereas other times the case is opposite.
Figure \ref{fig:blue-velocity} (c) and (g) show the time variation of the equivalent width of the Gaussian fitted to the residual for the one- and two-component fits, respectively.
As shown in Figure \ref{fig:blue-velocity} (g), the equivalent width of one of the two Gaussians was close to zero at the beginning and end of the time when the blue-shifted excess component was visible. Both equivalent widths were above a certain value in the intervening.
That is, the blue-shifted excess component initially appeared to be one component, then became two components, and finally appeared to be one component again. 

We assume that the prominence visibility for the Sun, where the prominence is considered the emission component of $\mathrm{H\alpha}$ and the filament is regarded the absorption component of $\mathrm{H\alpha}$ \citep{2014solar...11..1}, holds true here. Then, the fact that the blue-shifted excess component appears to be composed of two components can be interpreted in two ways:
\begin{enumerate}[(i)]
    \item Emission $+$ Emission: \\
    As shown in Figure \ref{fig:kaisyaku} (a), two prominences are present above the limb and each prominence is visible as an emission line in $\mathrm{H\alpha}$. This would be the situation when the prominence eruption occurred twice, or when the erupted prominence split in two while moving.
    \item Absorption $+$ Emission: \\
    As shown in Figure \ref{fig:kaisyaku} (b), the erupted prominence contains both the area above the limb and visible as an emission line in $\mathrm{H\alpha}$, and that inside the limb and visible as an absorption line in $\mathrm{H\alpha}$. When mixing of those emission and absorption components, two peaks seem to exist in the blue-shifted excess component.
    In this case, the width of the emission component must be broader than that of the absorption component to reproduce the observed spectrum. However, the physical interpretation for this situation is not well understood.
\end{enumerate}
Nevertheless, whether the premise of these interpretations can be applied to the present case remains unclear.
That is, unlike the Sun, filaments may be visible as the $\mathrm{H\alpha}$ emission components on K-type stars.
\citet{2022mnras...513..60586073} showed that for dM stars, thermal radiation from filaments dominates the source function over the scattering of the star's incident radiation so that even filaments can be considered emission line components of $\mathrm{H\alpha}$ through 1D NLTE modeling and cloud model formulation.
Given that our eruptive event occurred on a K-type star, which is cooler than the Sun, a filament may not necessarily be visible in the absorption component in $\mathrm{H\alpha}$, similar to that claimed by \citet{2022mnras...513..60586073} for M-type stars. 
Therefore, modeling and simulation of prominence/filament eruptions on K-type stars are required for a more advanced understanding of the two-peaked blue-shifted excess components.
\subsection{Comparison with other events} \label{sec:statistics}
We compared the fast prominence eruption observed on V1355 Orionis with other events regarding mass, velocity, and kinetic energy.
Figure \ref{fig:statistics} shows the (a) mass, (b) velocity, and (c) kinetic energy of the prominence eruptions and CMEs as a function of flare energy emitted in the GOES wavelength band (1-8 $\mathrm{\AA}$) and bolometric flare energy.
We used the equation (1) of \citet{2019apj...877..105} when converting the energy emitted in $\mathrm{H \alpha}$ to GOES X-ray flare energy: $L_{\mathrm{X}}=16 L_{\mathrm{H \alpha}}$. 
We also assumed the relationship: $L_{\mathrm{bol}} = 100 L_{\mathrm{X}}$ , which is shown to hold on solar flares by \citet{2012apj...759..71}. On the other hand, \citet{2015apj...809..79} shows $L_{\mathrm{bol}} = L_{\mathrm{X}} / 0.06$ for active stars. Therefore, the bolometric energy of the stellar flares in Figure \ref{fig:statistics} may be a bit smaller.
Note that for stars, only examples estimated from blue-shifted excess components of chromospheric lines are plotted in Figure \ref{fig:statistics}.
The lower and upper limits of the velocity of the prominence eruption on V1355 Orionis in Figure \ref{fig:statistics} (b) are the velocity obtained by fitting with two-component. The pink circle denotes the velocity obtained by fitting with one-component.
  
Figure \ref{fig:statistics} (a) indicates that the erupted prominence on V1355 Orionis has roughly the mass that expected from the scaling law of solar CMEs. This suggests that the prominence eruption on V1355 Orionis was caused by the same physical mechanism as the solar prominence eruptions/CMEs \citep[e.g.,][]{Kotani_2023}. Figure \ref{fig:statistics} (a) also shows that it is the largest prominence eruption observed by blue shift. These facts may provide important clues as to how large an event can be caused by the physical mechanism of solar prominence eruptions. Our observations may provide an opportunity to understand extreme eruptive events on stars. 

Figure \ref{fig:statistics} (b) shows that the prominence velocity on V1355 Orionis is indeed fast, but overwhelmingly below the theoretical upper limit of the velocity \citep{2016apjl...833..L8} estimated from the flare energy.
Therefore, the fast prominence eruption is physically possible to occur in association with a $10^{35} \: \mathrm{erg}$ class flare.

Figure \ref{fig:statistics} (c) indicates that kinetic energy corresponds roughly to the value predicted from the scaling law of the solar CMEs. In our calculation, the kinetic energy of the prominence is $4.5 \times 10^{33} \: \mathrm{erg} < K < 1.0 \times 10^{37} \: \mathrm{erg}$. 
As discussed in \citet{2019apj...877..105}, the kinetic energy of the prominence on other stars is smaller than that expected from the scaling law of solar CMEs. The possible reason for the discrepancy is that prominence eruptions generally have a lower velocity than CMEs in case of the Sun \citep{2021PASJ...73..44-65, 2022nature...6..241}, although it is also proposed that the suppression by the overlying large-scale magnetic field can contribute to small kinetic energies \citep{Alvarado-Gomez_2018}. 
For the blue-shift events except on V1355 Orionis, the kinetic energy tends to be smaller than the scaling law. 
However, we are not sure if the prominence of V1355 Orionis is below the scaling law due to the large indefiniteness of the kinetic energy.

Given these considerations, the prominence eruption on V1355 Orionis and solar prominence eruptions may have a common physical mechanism.
However, the large uncertainties in the mass estimate makes it difficult to compare them with the solar CME scaling law. 
For the mass estimation, we made various assumptions, as discussed in Section \ref{sec:Mass}, that may be incorrect.
A more accurate derivation of the prominence mass requires a simulation as performed in \citet{2022mnras...513..60586073}, as well as concerning the two components in Section \ref{sec:two-components}.

\section{Summary and Conclusion} \label{sec:conclusion}
We simultaneously performed spectroscopic observations in this study using the Seimei telescope and photometric observations using \textit{TESS} on the RS CVn-type star V1355 Orionis. We captured a superflare that releases $7.0 \times 10^{35} \: \mathrm{erg}$ and has the following characteristics:
\begin{enumerate}
    \item For the first 30 min after the flare started, a pronounced blue shift in the $\mathrm{H\alpha}$ emission line was observed confirming that the prominence eruption occurred in association with the flare.
    \item The velocity of the prominence eruption calculated from the blue-shift was up to $990 \: \mathrm{km \: s^{-1}}$ (one-component fitting) and $1690 \: \mathrm{km \: s^{-1}}$ (two-component fitting), that is, well above the escape velocity of $347 \: \mathrm{km \: s^{-1}}$.
    \item There seems two blue-shifted excess components with multiple possible interpretations.
    \item The mass of the prominence eruption is also one of the largest ever observed ($9.5 \times 10^{18} \: \mathrm{g} < M < 1.4 \times 10^{21} \: \mathrm{g}$), corresponding to the value expected from the flare energy-mass scaling law that holds for solar CMEs. However, the mass estimates make many uncertain assumptions and are highly indeterminate.
\end{enumerate}

In a very rare case, a prominence eruption at the velocity that greatly exceeds the escape velocity of the star was captured continuously at a high temporal resolution of 1 min simultaneously with a white light flare. 
The massive and fast prominence eruption detected in this study provide an important indicator of how large an eruption the physical mechanism of solar prominence eruptions can cause at most. 
Therefore, this will need to be investigated in the future with a larger sample of prominence eruptions in the energy range of $> 10^{35} \: \mathrm{erg}$.

As also discussed in \citet{2022mnras...513..60586073}, simply applying the empirical relation of solar prominence/filament visibility to star can involve many ambiguities. Further, modeling and simulation of prominence visibility on K-type stars are necessary to accurately interpret our data, especially the two-peaked blue-shifted excess component. This would also contribute to a more accurate derivation of the prominence mass.

\begin{acknowledgments}
The spectroscopic data used in this study were obtained through the program 20B-N-CN03 with the 3.8m Seimei telescope, which is located at Okayama Observatory of Kyoto University.
\textit{TESS} data were obtained from the MAST data archive at the Space Telescope Science Institute (STScI). 
All the \textit{TESS} data used in this paper can be found in MAST: \dataset[10.17909/ffwb-dg98]{http://dx.doi.org/10.17909/ffwb-dg98} 
Funding for the \textit{TESS} mission is provided by the NASA Explorer Program.
We thank T. Enoto (Kyoto University/RIKEN), H. Uchida, and T. Tsuru (Kyoto University) for their comments and discussions. 
We acknowledge the International Space Science Institute and the supported International Team 510: Solar Extreme Events: Setting Up a Paradigm (\url{https://www.issibern.ch/teams/solextremevent/}).
This research is supported by JSPS KAKENHI grant numbers 20K04032, 20H05643 (H.M.) 21J00106 (Y.N.), 21J00316 (K.N.) and 21H01131 (H.M., D.N., K.S.).
Y.N. was also supported by NASA ADAP award program number 80NSSC21K0632 (PI: Adam Kowalski).
\end{acknowledgments}

%

\vspace{5mm}
\facilities{Seimei \citep{2020PASJ...72..48} , \textit{TESS} \citep{2015...1..014003}}


\software{\texttt{astropy} \citep{astropy:2013, astropy:2018},
          \texttt{kools\_ifu\_red} \citep{2019PASJ...71..102},
          \texttt{IRAF} \citep{1986...627..733},
          \texttt{PyRAF} \citep{pyraf}
          }
          
\bibliography{sample631}{}
\bibliographystyle{aasjournal}



\end{document}